\documentclass[10pt]{bmc_article}

\usepackage{cite} % Make references as [1-4], not [1,2,3,4]
\usepackage{url}  % Formatting web addresses  
\usepackage{ifthen}  % Conditional 
\usepackage{multicol}   %Columns
\usepackage[utf8]{inputenc} %unicode support
\def\includegraphics{}

\newboolean{publ}

\newenvironment{bmcformat}{\begin{raggedright}\baselineskip20pt\sloppy\setboolean{publ}{false}}{\end{raggedright}\baselineskip20pt\sloppy}

\begin{document}
\begin{bmcformat}

%%%%%%%%%%%%%%%%%%%%%%%%%%%%%%%%%%%%%%%%%%%%%%
%%                                          %%
%% Enter the title of your article here     %%
%%                                          %%
%%%%%%%%%%%%%%%%%%%%%%%%%%%%%%%%%%%%%%%%%%%%%%

\title{Selection for smaller brains in Holocene human evolution}
 
%%%%%%%%%%%%%%%%%%%%%%%%%%%%%%%%%%%%%%%%%%%%%%
%%                                          %%
%% Enter the authors here                   %%
%%                                          %%
%% Ensure \and is entered between all but   %%
%% the last two authors. This will be       %%
%% replaced by a comma in the final article %%
%%                                          %%
%% Ensure there are no trailing spaces at   %% 
%% the ends of the lines                    %%     	
%%                                          %%
%%%%%%%%%%%%%%%%%%%%%%%%%%%%%%%%%%%%%%%%%%%%%%

\author{John Hawks\correspondingauthor$^{1}$%
       \email{John Hawks\correspondingauthor - jhawks@wisc.edu}%
}

%%%%%%%%%%%%%%%%%%%%%%%%%%%%%%%%%%%%%%%%%%%%%%
%%                                          %%
%% Enter the authors' addresses here        %%
%%                                          %%
%%%%%%%%%%%%%%%%%%%%%%%%%%%%%%%%%%%%%%%%%%%%%%

\address{%
    \iid(1)Department of Anthropology, University of Wisconsin--Madison, %
        5240 Sewall Social Sciences Building, 1180 Observatory Road, Madison WI 53706}%

\maketitle

%%%%%%%%%%%%%%%%%%%%%%%%%%%%%%%%%%%%%%%%%%%%%%
%%                                          %%
%% The Abstract begins here                 %%
%%                                          %%
%% The Section headings here are those for  %%
%% a Research article submitted to a        %%
%% BMC-Series journal.                      %%  
%%                                          %%
%% If your article is not of this type,     %%
%% then refer to the Instructions for       %%
%% authors on http://www.biomedcentral.com  %%
%% and change the section headings          %%
%% accordingly.                             %%   
%%                                          %%
%%%%%%%%%%%%%%%%%%%%%%%%%%%%%%%%%%%%%%%%%%%%%%

\begin{abstract}
        % Do not use inserted blank lines (ie \\) until main body of text.
        \paragraph*{Background:} Human populations during the last 10,000 years have undergone rapid decreases in average brain size as measured by endocranial volume or as estimated from linear measurements of the cranium. A null hypothesis to explain the evolution of brain size is that reductions result from genetic correlation of brain size with body mass or stature.
      
        \paragraph*{Results:} The absolute change of endocranial volume in the study samples was significantly greater than would be predicted from observed changes in body mass or stature. 

        \paragraph*{Conclusions:} The evolution of smaller brains in many recent human populations must have resulted from selection upon brain size itself or on other features more highly correlated with brain size than are gross body dimensions. This selection may have resulted from energetic or nutritional demands in Holocene populations, or to life history constraints on brain development. 
\end{abstract}

\ifthenelse{\boolean{publ}}{\begin{multicols}{2}}{}

%%%%%%%%%%%%%%%%%%%%%%%%%%%%%%%%%%%%%%%%%%%%%%
%%                                          %%
%% The Main Body begins here                %%
%%                                          %%
%% The Section headings here are those for  %%
%% a Research article submitted to a        %%
%% BMC-Series journal.                      %%  
%%                                          %%
%% If your article is not of this type,     %%
%% then refer to the instructions for       %%
%% authors on:                              %%
%% http://www.biomedcentral.com/info/authors%%
%% and change the section headings          %%
%% accordingly.                             %% 
%%                                          %%
%% See the Results and Discussion section   %%
%% for details on how to create sub-sections%%
%%                                          %%
%% use \cite{...} to cite references        %%
%%  \cite{koon} and                         %%
%%  \cite{oreg,khar,zvai,xjon,schn,pond}    %%
%%  \nocite{smith,marg,hunn,advi,koha,mouse}%%
%%                                          %%
%%%%%%%%%%%%%%%%%%%%%%%%%%%%%%%%%%%%%%%%%%%%%%

%%%%%%%%%%%%%%%%
%% Background %%
%%
\section*{Background}

An increase in brain size was one of the major trends of human evolution \cite{Ruff:1997, Lee:2003}. At the beginning of the Pleistocene, the average endocranial volume of fossil \emph{Homo} specimens was approximately 750 ml \cite{Rightmire:2004}. By 30,000 years ago, this average value had increased to nearly 1500 ml \cite{Ruff:1997, Lee:2003}. Much of this increase occurred within the period following 800,000 years ago \cite{Ruff:1997, Lee:2003}, during which mean endocranial volume in \emph{Homo} increased by approximately 70 ml per 100,000 years. This trend occurred in all regions of the Old World \cite{Lee:2003}, which may have included either a single \cite{accretion, Lee:2003} or multiple species of archaic \emph{Homo} \cite{Leigh:1992, Rightmire:2004}. \pb

Less well known is that the terminal Pleistocene and Holocene (ca. 30,000 years ago to present) witnessed a substantial \emph{decline} in endocranial volume \cite{Beals:1984, Henneberg:1988b, Ruff:1997}. This decrease occurred within modern \emph{Homo sapiens}, and has been observed in many parts of the world \cite{Beals:1984, Henneberg:1988b, Henneberg:Steyn:1993}. The scope of this decrease is remarkable: for example, within the past 10,000 years the average endocranial volume in European females reduced from a mean of 1502 ml to a recent value of 1241 ml \cite{Henneberg:1988b}. This decrease of approximately 240 ml in 10,000 years is nearly 36 times the rate of \emph{increase} during the previous 800,000 years. \pb

Brain size is related to body size both across higher taxa \cite{Jerison:1973} and within humans \cite{Holloway:brain-body:1980}. This suggests the hypothesis that changes in human brain size may result from changes in body size. For example, the larger brain size in early \emph{Homo} compared to \emph{Australopithecus} may reflect the simple expansion in body size from earlier hominids \cite{McHenry:2000}. This explanation cannot explain every change in brain size in humans: for example, the long increase in brain size during the Pleistocene did not coincide with increases in body size \cite{Rightmire:2004}. \pb

What about the reduction in brain size during the last 10,000 years---can it be explained by a reduction in the size of the body? Human body size, as measured from skeletal dimensions, did reduce during the past 30,000 years, at least in some populations \cite{Beals:1984, Henneberg:1988b, Ruff:1997, Ruff:2002}. This reduction influenced both mass and stature \cite{Henneberg:1988b, Ruff:1997, Ruff:2002}. A reduction in overall body size may have resulted from Late Pleistocene and Holocene subsistence strategies, which replaced close-contact ambush hunting of large mammals with projectile weapons, intensive collection of small animals, fish, and shellfish, and ultimately sedentary pastoralism and agriculture \cite{Frayer:1981}. Nutritional inadequacies and disease during the Holocene also may explain reductions in body size \cite{Armelagos:agriculture:1991}. Within Europe, where the trend has been most closely studied, body size rebounded within the past 1000 years as manifested by increases in stature \cite{Henneberg:1988b}. \pb

Several workers have suggested that recent reductions in brain size may have been caused by reductions in body size \cite{Beals:1984, Henneberg:1988b, Ruff:1997, Leach:2003}. A coincidence of reduction in both these measures would lend some support to that hypothesis. However, for a reduction in body size to be a sufficient explanation for reduction in brain size, it is not enough that the reductions occurred at the same time. Natural selection on one character (like body size) will affect a correlated character (like brain size) only to the extent that the two characters are heritable and are genetically correlated. Therefore, to test the hypothesis that selection on body size accounts for reductions in brain size in recent human evolution, we must consider the relationship and genetics of these characters within human populations. \pb

Here, I apply a quantitative genetic model to test the hypothesis that Holocene evolution of brain size may be explained by reductions in body size. The reasons for reduction in body size are unclear, so I consider both body mass and stature as candidates for the target of selection in recent populations. This is a very limited approach, constrained to published estimates of endocranial volume in archaeological populations and estimates of phenotypic correlations and heritability from samples of living humans. No attempt is made to correlate brain size and body size in the same samples of archaeological specimens, as such data are not available at present. Instead, I estimate the amount of body size change that would be necessary to explain the observed change in endocranial volume. This estimate is then assessed for credibility as applied to archaeological samples.  \pb

%%%%%%%%%%%%%%%%%%%%%%%%%%%%
%% Results and Discussion %%
%%
\section*{Results and Discussion}

  \subsection*{Body mass}

Body mass is related to brain size in humans with a phenotypic correlation of $r \approx 0.29$. The standard deviation of male body mass within recent human populations ranges around 11 kg, a value near the midpoint of within-sex variation in other primate species \cite{Smith:Jungers:1997}. Using these values along with the others listed in Table 1, selection on body mass would be expected to reduce the mean endocranial volume by 4.3 ml for each kilogram of reduction in body mass. \pb

The decline in body mass in human populations during the last 10,000 years has been estimated as less than 5 kg, or less than a 10 percent reduction in mass from a Late Upper Paleolithic mean of some 63 kg \cite{Ruff:1997}. A decline of 5 kg would predict a decrease in endocranial volume only around 22 ml. The observed decline in several regions (including Europe, China, Southern Africa, and Australia) is between 100 and 150 ml during the past 10,000 years. Therefore, the reduction in body mass would be expected to have decreased brain size by only one-fifth to one-seventh the observed decline. \pb

We can look at the inverse question: how much reduction in body mass would be required to cause a 150 ml reduction in endocranial volume? Using the same ratio (4.3 ml per kilogram body mass), the endocranial volume contrast would predict a reduction of 34 kg. This value is implausibly high, by more than a factor of five. \pb

The reduction of endocranial volume in these populations is not well explained by body mass according to equation 1. Selection for smaller mass is insufficient to account for reduction in brain size or vault dimensions. \pb

  \subsection*{Stature}

Applying equation 1 to the parameters for stature and its correlation to brain size, endocranial volume would be expected to change approximately 9.5 ml per centimeter change in stature. This value is less extreme than the reduction in body mass that would be necessary to achieve the same reduction in brain size. But the skeletal record is inconsistent with any great decrease of mean male stature, particularly during the post-Neolithic time period. \pb

Stature estimates exist for a broad sample of ancient European populations, showing approximate stasis in stature during the last 4000--6000 years. Over the same time period, the estimated endocranial volume declined slightly more than 100 ml in Europe from an estimated 1496 ml to 1391 ml. This decline cannot be explained by decreases in stature, because the stature did not change. Additionally, although these early samples are small, Mesolithic Europeans had larger endocranial volumes than Upper Paleolithic Europeans, across the same interval when they underwent a substantial decline in stature. That Mesolithic change in endocranial volume is in the opposite direction expected from the change in stature. \pb

Likewise, the femur lengths of foragers in Southern Africa showed no net decrease over the last 10,000 years. From 5500 to 2500 years ago, both femur length and femur head diameter declined in this region, but they rebounded within the last 2500 years \cite{Pfeiffer:2006}. Across the same 10,000-year time period, Henneberg and Steyn \cite{Henneberg:Steyn:1993} documented a decline in external and internal cranial module. The sample of LSA foragers (before 2000 years BP) had a mean external cranial module of 154.7, Iron Age (2000--200 years BP) had a mean of 149.6, while recent foragers had a mean of 150.3 --- roughly a standard deviation lower than the pre-2000 BP value. Under the hypothesis that change in endocranial volume is predicted by the change in stature, we should predict no net change in endocranial volume in this population. But the reduction in external module corresponds to a reduction in endocranial volume between 100 and 150 ml \cite{Henneberg:Steyn:1993}. However, the LSA sample in that study is very small ($n=12$) and temporally dispersed. \pb

Early Holocene populations in Australia have produced a substantial sample of crania, but postcrania from this time period are rare or poorly preserved \cite{Brown:Australia:1987}. The net change in endocranial volume, roughly 130 ml from the terminal Pleistocene to late Holocene skeletal sample \cite{Brown:1992} would predict a reduction in stature of 13 cm, if the brain size had changed only because of correlated changes in stature. That degree of stature reduction is not biologically impossible although it would be extreme. Further investigation of the evolution of body size in recent Australian hunter-gatherers may be necessary to answer the question. \pb

  \subsection*{Why did brain size reduce during the Holocene?}

The evidence suggests substantial reductions in brain size in some recent human populations, more than can be explained by correlated changes in body size. It is worth discussing two related points concerning the distribution and causes of this pattern of brain size evolution. \pb

First, was the change global or local in scope? The samples here cover several far-flung geographic areas, but they do not cover all regions of the world. Beals, Smith and Dodd \cite{Beals:1984} reviewed the global evidence for endocranial volume and showed a decline in the available terminal Pleistocene to Holocene skeletal sample. The Late Pleistocene skeletal sample was in that case strongly biased toward Europe, an area that in contemporary humans has a relatively large average endocranial volume. Thus, it was not obvious whether geographic differences in sampling might explain the reduction in endocranial volume noted in the study. This problem also characterizes the somewhat more course sampling by Ruff and colleagues \cite{Ruff:1997}. Here, the samples of endocranial volumes and body sizes are matched in region to the extent possible; they do represent probable evolutionary trends within these populations. But there are few other comparable sequences of skeletal samples, so it may not be possible to conclude strongly that the reduction in brain size generalizes outside these regions. \pb

A large series of crania from ancient Nubia covers the period from roughly 3400 years ago to 600 years ago \cite{Carlson:1976, VanGerven:1977}. Samples show a slight trend toward decrease in the major length, breadth and height measurements from Iron Age (Meroitic, external cranial module 145.2) to Medieval (Christian, external cranial module 143.9) times, but the intermediate series of crania (X-Group, external cranial module 147.1) is somewhat larger in these dimensions than either of the other groups. In this context it would be misleading to speak of a reduction in cranial vault size in this region. Across the same time interval, these samples show a substantial reduction in facial and dental measurements \cite{VanGerven:1977}. \pb

Second, given that the pattern is widespread if not global, how can we explain the reduction in brain size? Several hypotheses have been presented that may help to explain recent brain evolution. It is beyond the scope of this paper to test these hypotheses but here I review several of the adaptive and non-adaptive alternatives with some notes relating to the observed pattern. \pb

\begin{enumerate}

\item Chance. Genetic drift may be considered a null hypothesis for any slight morphological change. However, in the case of brain size evolution during the last 10,000 years, genetic drift is a markedly unlikely hypothesis. Endocranial volume changed by a standard deviation or more, rapidly and directionally, within some very numerous and growing post-agricultural populations. 

\item Plasticity. Somatic development in humans is plastic to some degree, depending on uterine and childhood nutritional and disease environments. This plasticity underlies most of the recent secular trend in body mass and stature. However, the brain size reaches 90 percent of its adult value very early in development and most of the variance in living populations is additive. This suggests that brain size may be less plastic than other components of body size. The pattern of decrease does not match stature or mass across the last several thousand years in these populations, suggesting that environmental effects were probably mediated by genetic factors. 

\item Climate. Beals, Smith and Dodd \cite{Beals:1984} presented correlations between endocranial volumes of populations and their local climate, as reflected by latitude or temperature. Smaller-brained populations live in warmer climates, and this relation cannot be explained entirely in terms of body size of contemporary populations. They proposed that post-glacial climate change may have favored smaller brains. However, if the link between climate and brain volume is not mediated through body mass (following Bergmann's rule), it is not obvious why climate should cause brain size reduction. 

\item Nutrition. The diets of early agriculturalists were nutritionally challenging in several ways: low in protein content, sometimes low in essential vitamins, and subject to fluctuating supply. The brain is an energetically expensive organ and nutritionally costly to develop. Smaller brains on balance should be advantageous under energetic or nutritional constraint, if they are functionally equivalent. Larger Holocene populations may have been selected for smaller brians for energetic reasons. 

\item Function. Smaller brains may have some functional implications, as white matter tracts are shorter and functional areas of the cortex may be more compact. Given the social and ecological changes of the Holocene, it is possible that a different mix of mental and cognitive functions was the target of selection. Despite the long Pleistocene history of human brain evolution, it would be fallacious to assume that larger brains were always adaptive in the context of cognitive changes.

\item Development. Although adult brain size is attained relatively early in development compared to adult body size, brain development continues during adolescence and early adulthood. It is possible that the life history evolution of recent humans has involved changes in the maturation schedule that would impact the ontogeny of brain maturation. If so, then the schedule of brain development after it attains adult size might have been constrained by earlier events, in such a way that faster development or smaller completed size was advantageous. 

\end{enumerate}

These hypotheses are not mutually exclusive. To assess them, it will be necessary to collect systematic data from a large sample of crania representing these and other regions of the world. This study represents only an early step toward understanding the cross-regional record of brain size evolution in the Holocene. \pb

Comparative data may also be useful to resolve these hypotheses. The decline of human endocranial volume during the last 10,000 years is paralleled most obviously by the reductions of brain size in domesticated animal species, including dogs, cattle and sheep, compared to their wild progenitors. Nutritional, developmental, and functional issues are all possible explanations for these parallel cases of brain size reduction. Humans are different in many ways from these domesticated species, but exhibit other parallel trends such as decreased skeletal robusticity.  \pb

At present, the literature presents a relative hodge-podge of estimates of endocranial volume, based on different original measurements. Estimates taken from the same method are compatible with each other, but it is not obvious that estimates based on different methods can be reconciled. It would be valuable to replace this mixture of measurements with a standard morphometric profile. The size of the endocranial cavity is interesting because of the developmental and energetic aspects of brains. But size is only one aspect of recent brain evolution. A full accounting of the shape of the cranial vault or endocast will be necessary to test hypotheses about why and how the brain reduced in size in these Holocene populations. \pb

%%%%%%%%%%%%%%%%%%%%%%
\section*{Conclusions}

The available skeletal samples show a reduction in endocranial volume or vault dimensions in Europe, southern Africa, China, and Australia during the Holocene. This reduction cannot be explained as an allometric consequence of reductions of body mass or stature in these populations. The large population numbers in these Holocene populations, particularly in post-agricultural Europe and China, rule out genetic drift as an explanation for smaller endocranial volume. This is likely to be true of African and Australian populations also, although the demographic information is less secure. Therefore, smaller endocranial volume was correlated with higher fitness during the recent evolution of these populations. Several hypotheses may explain the reduction of brain size in Holocene populations, and further work will be necessary to uncover the developmental and functional consequences of smaller brains. \pb

%%%%%%%%%%%%%%%%%%
\section*{Methods}
  \subsection*{Endocranial volume}
    
Studies of skeletal samples from different regions of the world are very consistent in finding reductions of endocranial volume during the last 10,000 years \cite{Beals:1984, Brown:Maeda:2004, Henneberg:1988b, Brown:1992, Henneberg:Steyn:1995, Henneberg:Steyn:1993, Schwidetsky:1977}. However, there are discrepancies among studies in the both the method of estimation and the time periods for which skeletal samples are available. These are listed in Table 1. \pb

\subsubsection*{Estimation methods}

The literature on brain size in archaeological specimens refers to several different measurements: 

\begin{enumerate}

\item Endocranial volume: directly measured by mustard seed, shot or water displacement of endocasts, or estimated from tomographic (CT) or magnetic resonance (MRI) methods. These different measurement methods can lead to systematically different results and so should not be combined without accounting for the measurement bias. The endocranial volume is larger than the brain volume (because of the intervening fluid and meningeal membranes). \pb

\item Brain weight: directly measured from cadavers or estimated from CT or MRI based on brain volume and estimated tissue density. \pb

Some notable large-sample studies of variation within contemporary human populations have examined brain weight \cite{Pakkenberg:1964}. Brain weight and endocranial volume are strongly correlated but not identical. The volume of the skull includes fluid and tissue components that are not included with cadaver brain weights, while different means of preservation of cadaver brains may inflate the variability of some brain weight datasets. The problems of brain weight measurement are not directly relevant to archaeological samples, where there are no brains to weigh. But brain weight remains important because of the present-day samples in which we can estimate the phenotypic correlation of brain and body size. Where possible, I have included present-day samples that include either endocranial volume or cranial measurements, for direct comparability with the archaeological samples. \pb

\item Cranial module: The external cranial module is the arithmetic mean of three external measurements of the skull: maximum length (glabella-opisthocranion), maximum breadth (euryon-euryon) and cranial height (basion-bregma). These external measurements include not only the brain but also the thickness of cranial bones. \pb

In some populations considered here, the thickness of cranial vault bones declined during the Holocene. This means that a decrease in the external module may be explained in part by a decrease in thickness, and some correction must be made to consider endocranial volume. The effect of thickness can be quite substantial; a decrease of 5 mm of thickness around a skull with an external module of 160 mm would increase its endocranial volume by around 180 ml. Where measurements of thickness are available, one approach is to subtract twice the vault thickness from the external module, resulting in an internal cranial module. This is the approach taken by Henneberg \cite{Henneberg:1988b}, for example, who reports both internal cranial module and resulting estimates of endocranial volume derived from regression on internal module. \pb

\end{enumerate}

The current paper uses the generic term ``brain size'' to refer to any of these estimation methods. Each of the four regions considered here is represented by at least one study that uses consistent estimation methods within the region. Even though different regions may be characterized by different methods of estimation, these differences should not bias the results \emph{within} each region. But when different regions produce a common result, it remains possible that the magnitude of changes may actually diverge from each other due to differences in estimation methods. \pb

One fundamental problem remains. Estimates of heritability and brain-body phenotypic correlation within human samples typically involve brain weight (for autopsy studies) or brain volume (for MRI or CT studies). Estimates from skeletal samples typically involve endocranial volume or cranial module. We cannot know that the heritability of the skeletal measures is equal to that of the soft-tissue measures.

\subsubsection*{Regions}

The literature includes sufficient data to consider the reduction of brain size in four regions of the world. \pb

The greatest temporal detail is available from Europe, reviewed by Henneberg \cite{Henneberg:1988b}. Samples of up to several thousand skulls have estimates of endocranial volume. The largest set of these are based on external measurements, corrected for average vault thickness. The literature also includes a substantial number of direct measurements of endocranial volume by seed or water displacement. Henneberg \cite{Henneberg:1988b} reports a Mesolithic mean endocranial volume for males of 1567 ml (based on internal cranial module of 144.1). This estimate is based on a relatively small sample of 35 individuals. For Neolithic and Eneolithic samples, with 1017 individuals, the mean endocranial volume estimate reduced to 1496 ml (internal cranial module 141.9), Bronze and Iron Age samples had a mean estimate of 1468 ml (internal cranial module 141.0), Roman period mean estimate 1452 ml (internal cranial module 140.5), and Early Middle Ages 1449 (internal cranial module 140.4). Late Middle Ages had a mean estimate 1418 (internal cranial module 139.4), and ``Modern Times'' (which comprises post-Medieval samples) corresponded to a mean estimate of 1391 ml (internal cranial module 138.5). Female samples across this time period exhibited a similar degree of size change; from a Neolithic mean of 1373 ml to 1210 ml in the ``Modern Times'' sample. \pb

Henneberg's study was notable for its discussion of the limitations of these data, which are compiled from many sources. The reliance on external dimensions does tend to increase the interstudy comparability of the values, but necessitates relying on regression predictions of endocranial volume, which necessarily involve some error. The overall change is substantial enough to overcome the plausible methodological inconsistencies, but it is appropriate to be cautious between time intervals (e.g., Early to Late Middle Ages) where the amount of change is minimal.

Endocranial volume in southern Africa was considered by Henneberg and Steyn \cite{Henneberg:Steyn:1993}, estimating from measurements of external and internal cranial module. The sample covers the time period after 30,000 radiocarbon years BP, however, the vast majority of specimens date to the last 2000 years. Henneberg and Steyn \cite{Henneberg:Steyn:1993} showed a statistically significant decline in both male and female crania, separated by morphological criteria. \pb

Much of this sample, together with a larger selection of archaeological crania, were included in a later study by Stynder and colleagues \cite{Stynder:2007} using morphometric methods. This study demonstrated an increase in craniofacial size during the last 4000 years, which appears to contradict the findings of Henneberg and Steyn \cite{Henneberg:Steyn:1993}. The resolution between these two results is twofold. Most obviously, Stynder and colleagues \cite{Stynder:2007} did not include landmarks that would indicate cranial breadth across the parietals, as these are not easily digitized. The breadth values are those showing the most consistent decreases in the sample studied by \cite{Henneberg:Steyn:1993}. Secondarily, Stynder et al. \cite{Stynder:2007} included facial measurements in their sample, so that the centroid size of crania was determined by both facial and vault dimensions. The allometric shape analyses in this paper demonstrated that larger centroid size was associated with allometric increase in the face and relative decrease in the vault. The implications of this allometry for the absolute vault dimensions are not clear, although the direct measurements indicate a reduction in vault size for the sample measured by Henneberg and Steyn \cite{Henneberg:Steyn:1993}. It would be valuable to look at these allometric questions comprehensively with both landmark and caliper measurements in the southern African sample. \pb

Brown and Maeda \cite{Brown:Maeda:2004, Brown:1992} reported on diachronic change of skeletal measurements in Holocene north China and Australia. They showed that the endocranial volume of males decreased from a mean of 1510 ml in early Neolithic (5500--6000 year old) samples down to 1400 ml in present-day Chinese. The change is consistent with a trend toward decrease across time intervals, despite relatively small sample sizes ($n=10$ to $n=20$ in the archaeological samples). Present-day Chinese people appear to vary in cranial size from north to south, possibly by more than 100 ml \cite{Brown:1992, Beals:1984}, and it is not obvious which samples of contemporary Chinese make the most relevant comparisons. So a decrease of 100 ml over the last 6000 years may either overstate or understate the actual change in endocranial volume in this population. \pb

Wu and colleagues \cite{Wu:2007} confirmed the trend toward smaller cranial size from Bronze Age to recent northern Chinese populations. The study included a much larger sample of crania than examined by Brown and Maeda \cite{Brown:Maeda:2004}, but endocranial volume itself was not measured. The length, breadth and height of the skull all underwent significant reductions from the Bronze Age, roughly 3000 years ago, to the present. \pb

Brown \cite{Brown:1992} presented a comparison of 19 male Australian crania from the terminal Pleistocene and 23 contemporary crania of Aboriginal Australians. The terminal Pleistocene sample stretches across a substantial range of dates, the earliest specimens possibly older than 30,000 years, to as little as 9000 for the large Coobool Creek sample. The Pleistocene people were larger in body size than recent Australians, and exhibit larger teeth and greater skeletal robusticity. The mean endocranial volume of the terminal Pleistocene males is 1405 ml; the recent mean is 1272 ml, for a decrease of just over 130 ml. \pb

In qualitative terms, the strongest documentation of the decline in endocranial volume is from Europe, due to both sample size and sample preservation. The other three skeletal samples show a comparable magnitude of decrease. In China, this decline occurred over roughly the same time interval as in Europe; in South Africa and Australia the reductions may have unfolded over a longer period of time. In all cases, the estimated reduction of endocranial volume was greater than 100 ml within males, roughly 7 percent of the mean. \pb

  \subsection*{Mass and stature}

Like brain size, stature and body mass provide challenges in the archaeological record. \pb

Mass is a parameter of fundamental biological interest, but it depends strongly on soft tissue body composition and is therefore estimated only with substantial error from skeletal samples. In a global survey of the Pleistocene human skeletal record, Ruff and colleagues \cite{Ruff:1997} estimated a mean body mass for Late Upper Paleolithic humans as 62.9 kg; this estimate was derived from 71 skeletal specimens, mostly from Europe. The ``living worldwide'' value cited in that study was 58.2 kg, a reduction of less than 5 kg from the Late Upper Paleolithic value, although the samples are geographically inconsistent. \pb
    
Stature should be a better proxy for body size in the archaeological record, because it exhibits less phenotypic plasticity and because it relates more directly to measurable skeletal quantities such as long bone lengths. This increases the geographic sample available to test hypotheses of temporal change, because either long bone lengths or stature estimates exist for Europe, Southern Africa, and China. \pb

Frayer \cite{Frayer:1981} reported an Upper Paleolithic male mean stature of 174 cm with a standard deviation of 9.4 cm. The Mesolithic male mean stature in that study was 165 cm with a standard deviation of 6.6 cm. The reduction in female stature values was concordant with the male values, with roughly half the number of sampled individuals. Maximum femur length reduced from 466 to 446 mm in male individuals between these time periods, with standard deviations of 38 and 29 mm, respectively. \pb

Henneberg \cite{Henneberg:1988b} lists a series of stature estimates from rural Poland since the 13th century. Both male and female statures were in approximate stasis over that time period, until the 19th century. Koepke and Baten \cite{Koepke:Baten:2005} put together a broader sample of anthropometric measures from across Europe during the last 2000 years, and also concluded that heights had been ``stagnant'' across that interval. Brief excursions of stature in some parts of Europe may nevertheless have occurred. Steckel \cite{Steckel:2004} collated a series of stature estimates from Northern European skeletal samples dating from the 9th to the 19th centuries. Across this region, the mean male stature declined from roughly 173.4 to a low of 166.2 cm during the 18th century, a reduction of 7 cm. That decline may have been presaged by an increase in the post-classical period suggested by the data of Koepke and Baten \cite{Koepke:Baten:2005}. Neither trend was noted in the samples considered by Frayer \cite{Frayer:1984} or Henneberg \cite{Henneberg:1988}. \pb

Sealy and Pfeiffer \cite{Sealy:2000} measured and performed stable isotope composition analysis of femora from the Cape region of South Africa, dating to the last 10,000 years. The male-attributed femora with measurable lengths in this study date to the period between 6000 and 1000 years ago. They show no significant decline in maximal length across this period. Femoral head diameter reduced slightly and significantly between the earlier male sample (before 4000 years ago) and later males (between 1000 and 4000 years ago). Pfeiffer and Sealy \cite{Pfeiffer:2006} revisited this sample and added evidence from more recnet skeletal individuals. The results showed that stature tended to rebound to a larger mean within the last 2000 years, roughly equal to the initial sample before 6000 years ago. Across this entire time period, the stature and mass of the archaeological population was within the range exhibited by present-day Khoisan peoples. \pb

The documentation of stature by long bone lengths is the best available source of data on body size in archaeological samples. Conservatively, we can conclude that the skeletal record documents a modest reduction of stature since the Upper Paleolithic in Europe, most of which had occurred by the Mesolithic. In Europe and China, the skeletal record is consistent with approximate stasis of stature during the last 5000 years, with some geographic and temporal excursions from the broad pattern. \pb

Body mass is unlikely to have changed is a very different pattern from stature. Fatness is poorly documented skeletally and is at present the largest component of variation in within-sex mass in industrial populations, but this varied much less substantially in pre-industrial peoples. \pb

  \subsection*{Quantitative genetic model}
  
For both body size parameters, the error of skeletal estimates is substantial. Therefore, here I adopt a very conservative test of the null hypothesis: (1) Determine the amount of change in body size that would minimally be required to explain the observed change in brain size; and (2) Evaluate whether that amount of change in body size is credible given the skeletal record. The skeletal record addresses point (2), but for point (1) we must turn to a quantitative genetic model relating the evolutionary dynamics of correlated characters. \pb

The allometry of brain and body size has been investigated extensively among both living and fossil organisms. From a quantitative genetic perspective, Lande \cite{Lande:1979} developed mathematical expectations for allometric change in the population mean of a single phenotypic character in response to selection on a correlated character. This change is given by Equation 2b in {Lande:1979}: 

\begin{equation}
\frac{\Delta\bar{z}_{i}}{\Delta\bar{z}_{b}} = \gamma_{ib}\frac{h_{i}\sigma_{i}}{h_{b}\sigma_{b}}
\label{eq:quant}
\end{equation}

$\Delta\bar{z}_{i}/\Delta\bar{z}_{b}$ indicates the change in the population mean $\bar{z}_{i}$ of one character (here, endocranial volume) with a correlated change in the mean $\bar{z}_{b}$ of a selected character (here, body size). The genetic correlation between the two characters is $ \gamma_{ib}$, while $h_{i}\sigma_{i}$ is the square root of the additive genetic variance of character $i$.  \pb

For this study, the null hypothesis is that brain volume should be predicted by  equation \ref{eq:quant}, given the parameter estimates and the change in body size. This is equivalent to the hypothesis that brain size has changed entirely due to its genetic correlation with body size. The parameters in equation \ref{eq:quant} have all been estimated in one or more contemporary human populations. \pb

It is important to note that parameter estimates may be conservative or nonconservative in their effects under the null hypothesis. The genetic correlation of the two traits must be less than 1. So measuring change in units of standard deviations, the null hypothesis predicts that brain size should change relatively less than body size. However, the absolute change must be considered relative to heritability and variance of the two phenotypic traits. Brain size should change \emph{more} relative to a given change in body size if: 

\begin{enumerate}
\item the genetic correlation of brain and body sizes is higher,
\item the heritability of brain size is higher, 
\item the phenotypic variation of brain size is higher, 
\item the heritability of body size is lower, or
\item the phenotypic variation of body size is lower.
\end{enumerate}

If the parameter estimates are in error in these directions, the test of the null hypothesis will be conservative to some degree --- that is, the null hypothesis will be accepted in cases where the true parameter values would lead to rejection. \pb

Estimates of heritability and variances are available for humans and for some other species of primates, both for brain volume and for body mass and stature. The availability of different estimates makes it possible to consider their consistency with each other and the likely effects of error. \pb

Mass and stature are considered separately as independent variables in the analysis. \pb

\subsubsection*{Brain size variation}

The skeletal samples above allow estimates of standard deviations for each sample. However, because of the limited sizes of archaeological samples, these estimates of variability may either overstate or understate the variation of ancient populations. \pb

There is substantial sexual dimorphism of both brain and body size in humans. The simplest way to correct for variation due to sex is to consider males and females separately. All estimates of parameter values in living humans are reported from male- or female-specific samples. Archaeological samples often permit assessment of individual sex, although there is necessarily some error in these assessments. Where possible, this study reports values for males, and assumes that variation is distributed like that of males in living human populations. \pb

Additionally, phenotypic estimates in humans may include confounding age effects. A few cited studies use age-controlled samples, but many rely on postmortem measures in samples with a broad range of age-at-death. Archaeological samples always include age-related variability, although this is likely distributed differently than in many surveys of living humans. \pb

Peper and colleagues \cite{Peper:2007} reviewed heritability estimates for total and regional brain volume based on MRI studies of twins. Most studies have yielded high estimates for the heritability of total brain volume, ranging from 0.97 \cite{Pennington:2000}, 0.94 \cite{Bartley:1997}, 0.90 \cite{Baare:2001} and 0.89 \cite{Wallace:2006}. One outlier study reported a lower estimate of heritability (0.66), but this came from a sample of only 10 MZ and 10 DZ twin pairs \cite{Wright:brain:2002}. In the current study, the use of a high estimate of heritability will tend to bias the result toward accepting the null hypothesis, since a more heritable character will be expected to change more under the effect of correlation with body size. \pb

\subsubsection*{Brain-body genetic correlation}

The genetic correlation between brain size and body size is not known for humans. However, the phenotypic correlations between brain volume or mass and body mass or stature have been extensively studied. The largest sample of these metrics was published from Danish autopsies by Pakkenberg and Voight \cite{Pakkenberg:1964}. Holloway \cite{Holloway:brain-body:1980} computed correlations between brain mass, stature and body mass in this dataset; these are reported in Table 1. \pb

Ankney \cite{Ankney:1992}, using the data from Ho et al. \cite{Ho:1980}, reports phenotypic correlations between brain mass and stature as $r=0.20$ for white males and $r=0.24$ for white females, $r=0.20$ for black males and $r=0.15$ for black females. These values are lower than those computed from the Danish data. Both sets of estimates should be regarded as underestimates because of the confounding effect of age variation in the sample. On the other hand, these are phenotypic correlations, and the genetic correlation may be lower than the phenotpic values due to effects induced by the environment or gene-environment interactions. Here, I employ the higher reported estimates of correlations because they have a conservative effect on the hypothesis test: A higher correlation predicts a more substantial change in brain size. \pb

\subsubsection*{Parameter values in nonhuman primates}

Estimates of brain-body correlations and heritabilities in humans have mostly been taken in European or American population samples. These estimates may therefore be biased dietary Westernization and concomitant changes in body mass index. To address this possibility, we can consider these relationships in non-human primates.  \pb

Rogers and colleagues \cite{Rogers:Jeffrey:2007} measured brain volume and body mass in captive free-ranging baboons (\emph{Papio hamadryas}) with known pedigrees. They found brain-body phenotypic correlation of $r=0.29$ ($r^2=0.086$) for males and $r=0.16$ ($r^2=0.026$) for females. The heritability of brain volume was estimated as 0.52. The heritability of body mass in this captive population was previously estimated as 0.50 \cite{Jaquish:1997}. \pb

Falk and colleagues \cite{Falk:1999} found phenotypic correlations in rhesus macaques (\emph{Macaca mulatta}) between brain volume and body mass to be $r=0.54$ for males and $r=0.40$ for females.  \pb

Stature is not strictly comparable between humans and other primates, because of the obvious difference in locomotor anatomy. \pb

These comparisons allow several conclusions: 

\begin{enumerate}
\item The heritability of body mass is approximately the same in humans as in other primates. 
\item Heritability of brain size in humans is substantially higher than reported in other primates. Using a high estimate should bias against rejection of the null hypothesis. 
\item The phenotypic correlations between brain size and mass in these primates are within the range reported for humans. 
\end{enumerate}

Thus, as near as possible, using the human values for these parameter estimates will provide an appropriate test of the null hypothesis, that changes in brain size were caused by changes in body size in recent human populations. \pb

%%%%%%%%%%%%%%%%%%%%%%%%%%%%%%%%
\section*{Authors contributions}
    The corresponding author conducted the research and wrote the manuscript.

%%%%%%%%%%%%%%%%%%%%%%%%%%%
\section*{Acknowledgements}
  \ifthenelse{\boolean{publ}}{\small}{}
  The work was supported by the Graduate School of the University of Wisconsin--Madison.

%%%%%%%%%%%%%%%%%%%%%%%%%%%%%%%%%%%%%%%%%%%%%%%%%%%%%%%%%%%%%
%%                  The Bibliography                       %%
%%                                                         %%              
%%  Bmc_article.bst  will be used to                       %%
%%  create a .BBL file for submission, which includes      %%
%%  XML structured for BMC.                                %%
%%                                                         %%
%%                                                         %%
%%  Note that the displayed Bibliography will not          %% 
%%  necessarily be rendered by Latex exactly as specified  %%
%%  in the online Instructions for Authors.                %% 
%%                                                         %%
%%%%%%%%%%%%%%%%%%%%%%%%%%%%%%%%%%%%%%%%%%%%%%%%%%%%%%%%%%%%%

{\ifthenelse{\boolean{publ}}{\footnotesize}{\small}
 \bibliographystyle{bmc_article}  % Style BST file
  \bibliography{refs,refs-2} }     % Bibliography file (usually '*.bib' ) 

%%%%%%%%%%%

\ifthenelse{\boolean{publ}}{\end{multicols}}{}

%%%%%%%%%%%%%%%%%%%%%%%%%%%%%%%%%%%
%%                               %%
%% Figures                       %%
%%                               %%
%% NB: this is for captions and  %%
%% Titles. All graphics must be  %%
%% submitted separately and NOT  %%
%% included in the Tex document  %%
%%                               %%
%%%%%%%%%%%%%%%%%%%%%%%%%%%%%%%%%%%

%%
%% Do not use \listoffigures as most will included as separate files

%\section*{Figures}
%  \subsection*{Figure 1 - Sample figure title}
%%      A short description of the figure content
%      should go here.

%  \subsection*{Figure 2 - Sample figure title}
%      Figure legend text.

%%%%%%%%%%%%%%%%%%%%%%%%%%%%%%%%%%%
%%                               %%
%% Tables                        %%
%%                               %%
%%%%%%%%%%%%%%%%%%%%%%%%%%%%%%%%%%%

%% Use of \listoftables is discouraged.
%%
\section*{Tables}
  \subsection*{Table 1 - Estimates of quantitative genetic parameters}
    Correlations and heritabilities of human brain and body dimensions used in this study. Values are from combined-sex samples. $^{\textrm{a}}$ Based on a range of estimates from several countries. $^{\textrm{b}}$ Age-matched sample. $^{\textrm{c}}$ Correlations taken from \cite{Holloway:brain-body:1980} based on original data from \cite{Pakkenberg:1964} and other sources cited therein. \par \mbox{}
    \par
    \mbox{
\begin{tabular}{lll}
Parameter & Value & Source\\
\hline
Brain volume heritability ($h^2$) & 0.94 &  \cite{Bartley:1997} \\
Stature heritability ($h^2$) & 0.80 $^{\textrm{a}}$ & \cite{Silventoinen:2003} \\
Body mass heritability ($h^2$) & 0.52 $^{\textrm{b}}$ & \cite{Mathias:2003} \\
Brain size--stature correlation & 0.47 $^{\textrm{c}}$ & \cite{Holloway:brain-body:1980}  \\
Brain size--body mass correlation & 0.29 & \cite{Holloway:brain-body:1980} \\
\end{tabular}
      }

%%%%%%%%%%%%%%%%%%%%%%%%%%%%%%%%%%%
%%                               %%
%% Additional Files              %%
%%                               %%
%%%%%%%%%%%%%%%%%%%%%%%%%%%%%%%%%%%

%\section*{Additional Files}
%  \subsection*{Additional file 1 --- Sample additional file title}
%    Additional file descriptions text (including details of how to
%    view the file, if it is in a non-standard format or the file extension).  %This might
%    refer to a multi-page table or a figure.
%
%  \subsection*{Additional file 2 --- Sample additional file title}
%    Additional file descriptions text.

\end{bmcformat}
\end{document}